\documentclass{iopart}
\usepackage{graphicx}

\newcommand{\bk}{{\bf k}}
\renewcommand{\Im}{\mathop{\mathrm{Im}}\nolimits}

\advance\textheight by \baselineskip

\begin{document}
\article[Pressure-induced ETTs in low dimensional
   superconductors]{EHPRG Award Lecture}{Pressure-induced electronic topological
   transitions in low dimensional superconductors}
\author{G. G. N. Angilella}
\address{Dipartimento di Fisica e Astronomia, Universit\`a di
   Catania,\\
and Istituto Nazionale per la Fisica della Materia, UdR Catania,\\
64, Via S. Sofia, I-95123 Catania, Italy}
\ead{Giuseppe.Angilella@ct.infn.it}

\begin{abstract}
The high-$T_c$ cuprate superconductors are characterized by a
   quasi-two-dimensional layered structure where most of the physics relevant
   for high-$T_c$ superconductivity is believed to take place.
In such compounds, the unusual dependence of critical temperature
   $T_c$ on external pressure results from the combination of the
   nonmonotonic dependence of $T_c$ on hole doping or hole-doping
   distribution among inequivalent layers, and from an ``intrinsic''
   contribution.
After reviewing our work on the interplay among $T_c$, 
   hole content, and pressure in the bilayered and multilayered
   cuprate superconductors, we will discuss how the proximity to an
   electronic topological transition (ETT) may give a microscopic
   justification of the ``intrinsic'' pressure dependence of $T_c$ in
   the cuprates.
An ETT takes place when some external agent, such as doping,
   hydrostatic pressure, or anisotropic strain, modifies the topology
   of the Fermi surface of an electronic system.
As a function of the critical parameter $z$, measuring the distance of
   the chemical potential from the ETT, we recover a nonmonotonic
   behaviour of the superconducting gap at $T=0$, regardless of the
   pairing symmetry of the order parameter.
This is in agreement with the trend observed for $T_c$ as a function
   of pressure and other material specific quantities in several
   high-$T_c$ cuprates and other low dimensional superconductors.
In the case of epitaxially strained cuprate thin films, we argue that
   an ETT can be driven by a strain-induced 
   modification of the in-plane band structure, at constant hole content, at
   variance with a doping-induced ETT, as is usually assumed.
We also find that an increase of the in-plane anisotropy enhances the
   effect of fluctuations above $T_c$ on the normal-state transport
   properties, which is a fingerprint of quantum criticality at $T=0$.
\end{abstract}


\maketitle

\section{Introduction}

High pressure research plays an important role in the study of
   superconducting materials.
A surprisingly large number of elemental solids is found to be
   superconducting under extreme conditions \cite{Ashcroft:02}, with
   the recent addition of nonmagnetic iron, with a critical
   temperature $T_c \sim 2$~K at pressures $P=15-30$~GPa \cite{Shimizu:01}, and
   possibly of lithium ($T_c = 20$~K, $P=48$~GPa) \cite{Shimizu:02a,Christensen:01a}.
The case of the ``simple'' alkali metals under pressure is actually
   quite interesting by itself.
Following earlier theoretical suggestions of an electronic \cite{Siringo:97}
   or pairing \cite{Neaton:99} instability, recent experimental
   findings actually indicate a metal-insulator transition into a
   reduced-symmetry phase at $P=45$~GPa
   \cite{Hanfland:99,Hanfland:00,Christensen:01}.
Our recent proposal is that Friedel oscillations in the pair potential
   may justify phase oscillations between a symmetric
   phase and a low-symmetry dimerized structure as a function of
   pressure, with re-entrant metallicity occurring at higher pressures
   \cite{Angilella:02e}.

Here, we will focus on the pressure effects on the superconducting
   properties of the high-$T_c$ cuprates.
These materials are characterized by CuO$_2$ layers, where most of the
   physics relevant for high-$T_c$ superconductivity
   is believed to take place \cite{Anderson:97}.
As a consequence of their reduced dimensionality, hydrostatic pressure
   as well as uniaxial strain are expected to influence remarkably the
   superconducting properties of the cuprates.
This is usually indicated by the magnitude and sign of the pressure
   derivative of $T_c$, $dT_c /dP$.
While in the case of `conventional', phonon-mediated superconductors
   one almost invariably has $dT_c /dP <0$, mainly as a result of
   lattice stiffening \cite{Schilling:01,MgB2}, the pressure dependence of
   $T_c$ in the cuprates is usually nonmonotonic, with a sign-changing $dT_c /dP$, and
   $T_c$ may even display kinks as a function of pressure
   \cite{Wijngaarden:99}.

Such a behaviour is usually related to an interplay among $T_c$, hole doping $n$,
   and pressure.
Since hole doping is itself a function of pressure, one may generally
   write \cite{Schilling:01}:
\begin{equation}
\frac{dT_c}{dP} = \frac{\partial T_c}{\partial P} + \frac{\partial
   T_c}{\partial n} \frac{dn}{dP} ,
\label{eq:Schilling}
\end{equation}
where $T_c$ at ambient pressure follows a universal
   bell-shaped curve as a function of doping \cite{Zhang:93}.
Equation~(\ref{eq:Schilling}) shows that the actual value and sign of
   the total derivative $dT_c /dP$ is given by the competition of a
   hole-driven contribution, and an ``intrinsic'' contribution,
   $\partial T_c /\partial P$.
The theoretical implications of such a pressure-dependent $T_c$, and
   in particular of a nonzero ``intrinsic'' contribution $\partial T_c
   /\partial P$, can help understanding the pairing mechanism of these
   unconventional superconductors.

In this Lecture, we will review our results concerning the interplay
   expressed by Equation~(\ref{eq:Schilling}), both for a bilayered
   cuprate \cite{Angilella:96} and for multilayered cuprates
   \cite{Angilella:99b}.
In the case of a bilayered cuprate (Section~\ref{sec:single}), after
   recovering the universal nonmonotonic dependence of $T_c$ on hole
   doping, we will extract a phenomenological estimate of the
   ``intrinsic'' contribution $\partial T_c /\partial P$ by comparison
   with available experimental data for Bi2212.
In the case of multilayered cuprates (Section~\ref{sec:multi}), we
   will emphasize the role of a nonuniform hole-content distribution
   among inequivalent CuO$_2$ layers.
Two possible sources of inequivalence will be considered, namely a
   different proximity to the charge-reservoir layers, and the
   different number of adjacent CuO$_2$ layers to which a given layer
   may be coupled by means of interlayer pair tunneling (ILT)
   \cite{Chakravarty:93,Angilella:99,Angilella:00}.
Later in Section~\ref{sec:ETT}, we will discuss the microscopic origin
   of the ``intrinsic'' pressure dependence of $T_c$ as due to the
   proximity to an electronic topological transition (ETT)
   \cite{Angilella:01,Angilella:02d,Angilella:03g}, which takes place
   when some external agent (such as pressure) modifies the topology
   of the Fermi surface of an electronic system.
We will eventually summarize in Section~\ref{sec:summary}.

\section{Interplay among $T_c$, hole doping, and pressure: bilayered cuprates}
\label{sec:single}

In Reference~\cite{Angilella:96}, we have analyzed a model Hamiltonian
   describing tightly bound interacting fermions in a bilayer complex:
\begin{equation}
H = \sum_{\bk\sigma} \xi_\bk c^\dag_{\bk\sigma} c_{\bk\sigma} +
   \frac{1}{N} \sum_{\bk\bk^\prime} V_{\bk\bk^\prime}
   c_{\bk\uparrow}^\dag c_{-\bk\downarrow}^\dag c_{-\bk^\prime
   \downarrow} c_{\bk^\prime \uparrow} ,
\label{eq:Hamiltonian}
\end{equation}
where $c_{\bk\sigma}^\dag$ ($c_{\bk\sigma}$) creates (destroys) a
   fermion with spin projection $\sigma$ along a specified direction,
   wave-vector $\bk$ belonging to the first Brillouin zone (1BZ) of a tetragonal
   lattice containing $N$ $\bk$-points, and band dispersion $\xi_\bk = \epsilon_\bk
   -\mu$, measured relative to the chemical potential $\mu$.
As for the tight-binding dispersion relation, we assumed the form
\begin{equation}
\epsilon_\bk = -2t [\cos (k_x a) + \cos (k_y a)] + 4 t^\prime \cos (k_x a)
   \cos (k_y a) -2 t_z \cos (k_z c),
\label{eq:disp}
\end{equation}
which retains in-plane nearest neighbour ($t$) and next-nearest
   neighbour ($t^\prime$) hopping, as well as nearest neighbour interlayer
   single-particle hopping ($t_z \ll t$).
The interaction term in Equation~(\ref{eq:Hamiltonian}) may be
   expanded over the different symmetry channels allowed by the
   $C_{4v}$ symmetry of the lattice.
Assuming the interaction kernel to be separable, one has $V_{\bk\bk^\prime} = \sum_h
   \lambda_h g_\bk^{(h)} g_{\bk^\prime}^{(h)}$, where $\lambda_h$ are
   phenomenological coupling constants, $g_\bk^{(0)} = 1$
   and $g_\bk^{(1)} = \frac{1}{2} [\cos (k_x a ) + \cos (k_y a)]$
   correspond to $s$-wave symmetry, and $g_\bk^{(2)} = \frac{1}{2}
   [\cos (k_x a ) - \cos (k_y a)]$ corresponds to $d$-wave symmetry
   \cite{Angilella:96,Angilella:99}.

A mean-field analysis of Equation~(\ref{eq:Hamiltonian}) as a function
   of hole content allowed us to recover the universal, nonmonotonic
   doping dependence of $T_c$ at ambient pressure \cite{Zhang:93}.
This seems to be a generic feature of a dispersion relation like
   Equation~(\ref{eq:disp}), as we will comment later on in
   Section~\ref{sec:ETT}.
While superconductivity sets in with a definite symmetry (either $s$-
   or $d$-wave) at $T=T_c$, with $d$-wave prevailing for hole contents
   larger than or equal to optimal doping, symmetry mixing is allowed at
   low temperature and intermediate dopings
   \cite{Angilella:96,Angilella:99}.
The role of the symmetry of the order parameter in determining the
   scale of $T_c$ in both high-$T_c$ superconductors and heavy fermion
   compounds has been discussed in
   References~\cite{Angilella:00b,Angilella:01a,Angilella:02g}.

\begin{figure}[t]
\centering
\includegraphics[height=0.45\columnwidth,angle=-90]{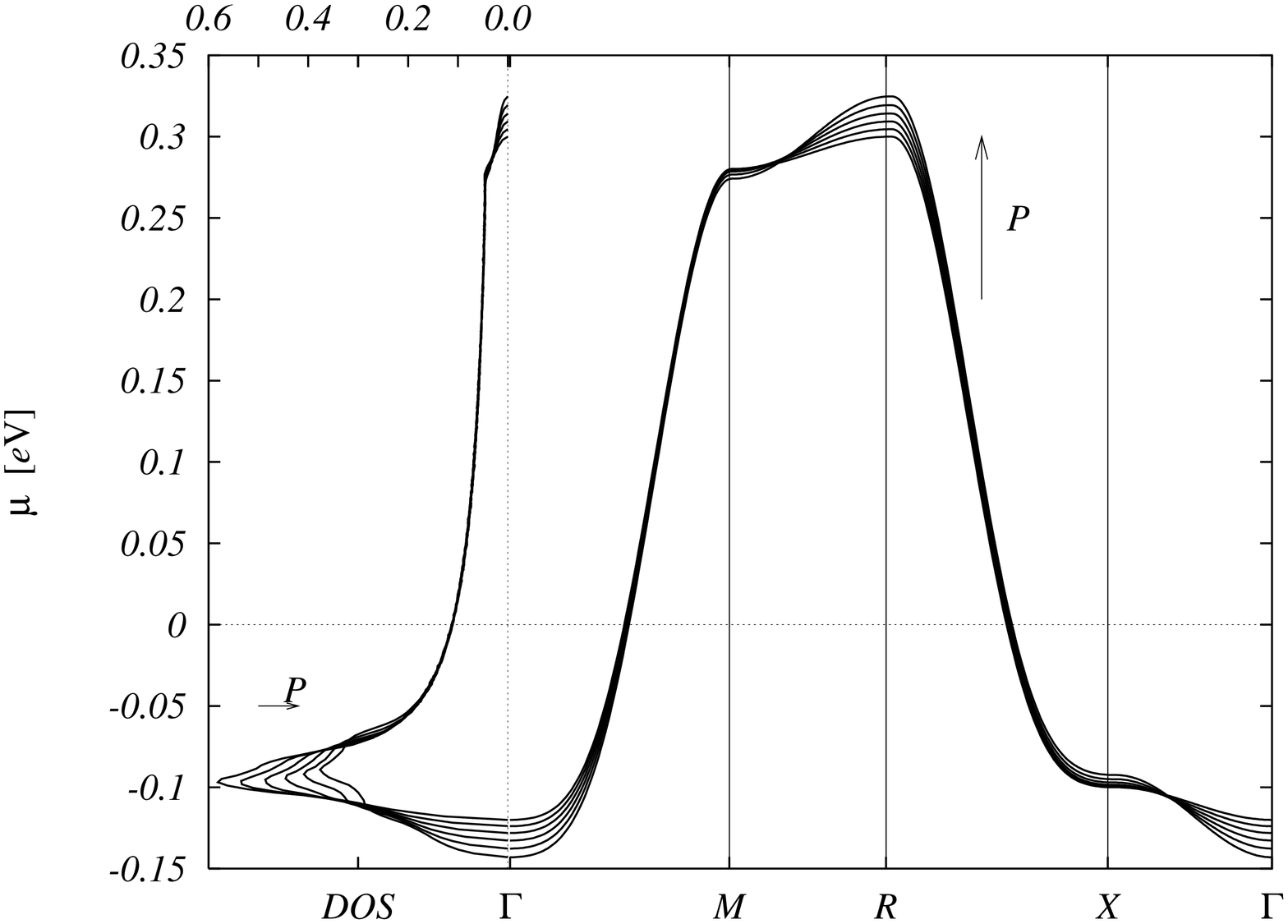}
\includegraphics[height=0.45\columnwidth,angle=-90]{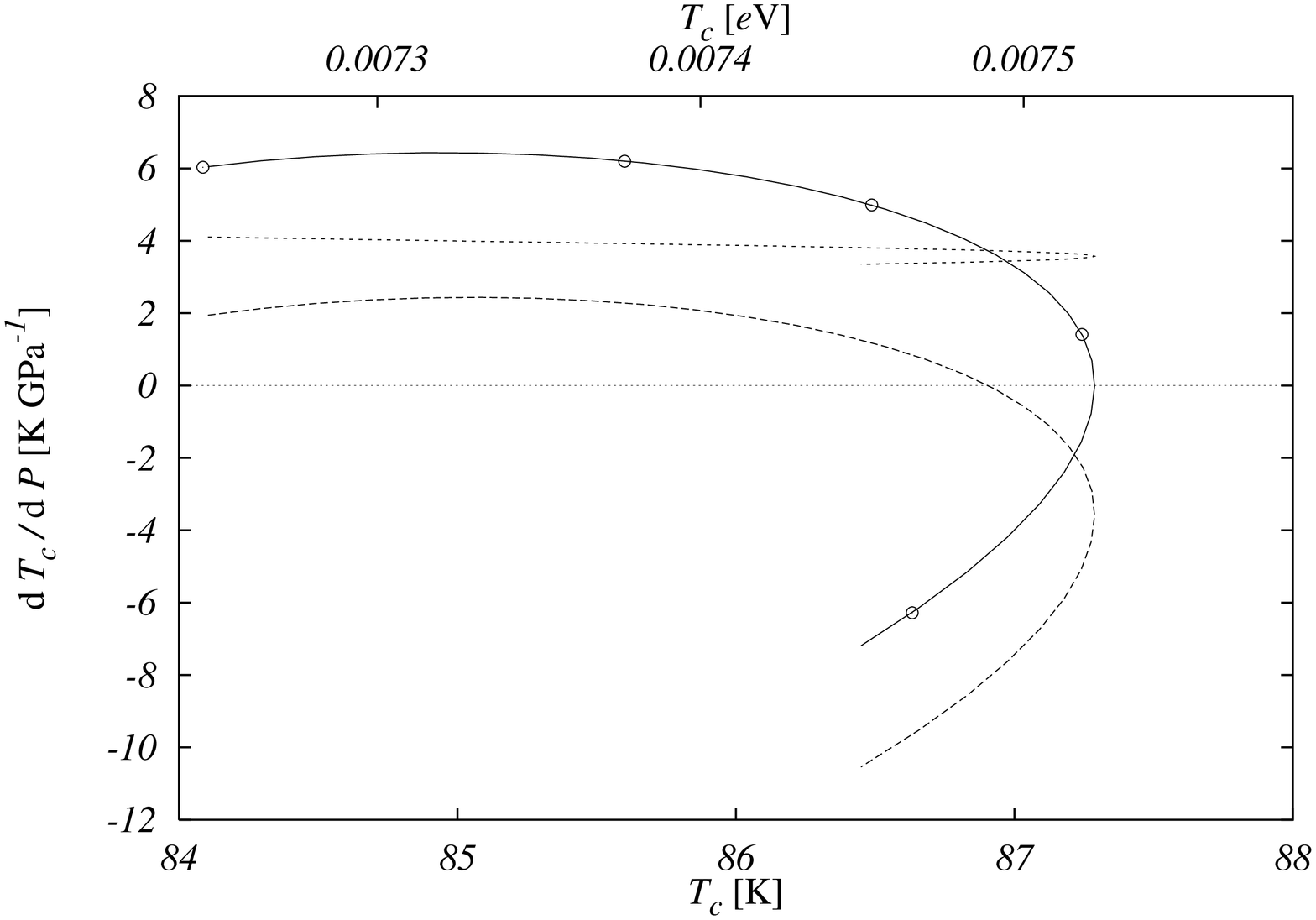}
\caption{{\sl Left:} Band dispersion along a symmetry contour of the
   1BZ, Equation~(\protect\ref{eq:disp}), and density of states as a
   function of pressure $P=0-20$~GPa.
{\sl Right:} Pressure derivatives of $T_c$ for $P=0-1.6$~GPa.
Solid line: best fit to the experimental data of Huang
   \emph{et al.} \protect\cite{Huang:93};
Dashed line: hole induced contribution;
Dotted line: our estimate for the ``intrinsic'' contribution,
   $\partial T_c /\partial P$ [Equation~(\protect\ref{eq:Schilling})].
Redrawn after Reference~\protect\cite{Angilella:96}.}
\label{fig:band}
\end{figure}

Such an analysis has been generalized for a nonzero applied pressure
   $P$.
The effect of pressure is mainly that of decreasing the lattice
   constants $a$ and $c$, thus varying the band parameters $t$, $t^\prime$, $t_z$,
   and of changing the hole content $n$, thus varying the chemical
   potential $\mu$.
While the pressure dependence of the lattice parameters may be
   extracted from the available components of the isothermal
   compressibility tensor $\kappa_i$ ($i=x,y,z$), the pressure
   dependence of the overall hole content can be derived from
   available data for the Hall resistance as a function of pressure
   \cite{Huang:93}.
Application of hydrostatic pressure results in widening of the
   electronic band, and in a decrease of the density of states, due to
   a reduction of the would-be Van~Hove singularity
   (Fig.~\ref{fig:band}).

By comparison with the available pressure dependence of $T_c$ in
   Bi2212 \cite{Huang:93}, we could extract a phenomenological
   estimate of the ``intrinsic'' contribution $\partial T_c /\partial
   P$ to the total pressure derivative of $T_c$ in
   Equation~(\protect\ref{eq:Schilling}) (Fig.~\ref{fig:band}).
Such a contribution turns out to be nonnegligible, especially at lower
   pressures, and justifies a sign changing overall $dT_c /dP$ with
   increasing pressure.

\section{Interplay among $T_c$, hole distribution, and pressure: multilayered cuprates}
\label{sec:multi}

In order to study the case of multilayered cuprate compounds, such as
   the homologous series Bi$_2$Sr$_2$Ca$_{n-1}$Cu$_n$O$_{2n+4+y}$
   (Bi-2:2:$(n-1)$:$n$, whereof Bi2212 is the bilayered instance, with
   $n=2$), Tl$_m$Ba$_2$Ca$_{n-1}$Cu$_n$O$_{2(n+1)+m+y}$
   (Tl-$m$:2:$(n-1)$:$n$, with $m=1,2$), or
   HgBa$_2$Ca$_{n-1}$Cu$_n$O$_{2n+2+y}$ (Hg-1:2:$(n-1)$:$n$), we have
   to generalize the above model in order to take into account for the
   inequivalence among different layers.
In Reference~\cite{Angilella:99b} we have considered basically two
   sources of inequivalence among different layers.

First of all, a different proximity to the `charge reservoir' layers may induce a
   nonuniform hole-content \emph{distribution} between inner and outer
   CuO$_2$ layers.
This may be taken into account by assuming a model electronic band in
   each layer
   as in Equation~(\ref{eq:disp}), but now with a chemical potential $\mu^\ell$
   depending on layer index $\ell$.

In addition to that, inner CuO$_2$ layers are expected to be coupled more
   intensely to adjacent layers than outer layers.
An intensely debated proposal for the origin of interlayer coupling in
   the cuprates is the interlayer pair tunneling (ILT) model
   \cite{Chakravarty:93,Anderson:97,Angilella:99,Angilella:00}.

The ILT mechanism of high-$T_c$ superconductivity has been proposed as
   a possible explanation for the observed high values of
   $T_c$ in the layered cuprates, as well as a number of other more
   difficult but related aspects of their complex phenomenology
   \cite{Anderson:97}.
The ILT mechanism relies on the fact that single-particle coherent
   tunneling between adjacent layers is suppressed, due to the
   so-called Anderson orthogonality catastrophe \cite{Anderson:97}.
Such an assumption amounts to set $t_z = 0$ in
   Equation~(\ref{eq:disp}) above.
On the other hand, coherent tunneling of Cooper \emph{pairs} does not
   suffer from such restrictions, and thus enters the total
   Hamiltonian as a second order effect in the single particle hopping
   matrix element $t_\perp (\bk)$ \cite{Chakravarty:93}.
While within conventional models the superconducting state is accessed
   by lowering the potential energy of the electronic system, the main
   aspect of the ILT mechanism is that Josephson tunnelling of Cooper
   pairs between adjacent CuO$_2$ layers allows the system to access
   the normal-state frustrated $c$-axis kinetic energy.
However, the ILT mechanism has been recently deeply reconsidered, due
   to its apparent inconsistency with the observed $c$-axis
   electrodynamics \cite{Leggett:98,Moler:98,Anderson:98}.

In the presence of ILT between adjacent layers, the interaction term
   in Equation~(\ref{eq:Hamiltonian}) should be modified as:
\begin{equation}
\frac{1}{N} V_{\bk\bk^\prime} \mapsto \,\,\,
\tilde{V}_{\bk\bk^\prime}^{\ell\ell^\prime} = \frac{1}{N}
   V_{\bk\bk^\prime} \delta_{\ell\ell^\prime} - T_J (\bk)
   \delta_{\bk\bk^\prime} (1-\delta_{\ell\ell^\prime}
   )\theta(1-|\ell-\ell^\prime |),
\end{equation}
where $\theta(\tau)$ is the usual Heaviside step function,
   $V_{\bk\bk^\prime}$ is the in-plane potential, now allowing for
   $d$-wave pairing only, and $T_J (\bk) =
   t_\perp^2 (\bk) /t$ is the ILT effective interaction
   \cite{Chakravarty:93,Angilella:99b}.
A standard mean-field treatment of the ensuing Hamiltonian yields a
   system of \emph{coupled} nonlinear equations for the gap function
   $\Delta_\bk^\ell$ on layer $\ell$ ($\ell=1,\ldots n$).
Therefore, linearization of these equations close to $T_c$ admit in
   principle $n$ solutions, $T_c = T_c^{(\ell)}$, say, corresponding
   to an incipiently nonzero gap $\Delta_\bk^\ell$ on layer $\ell$.
The `physical' solution for the critical temperature $T_c$ is then
   simply $\displaystyle T_c = \max_\ell  T_c^{(\ell)}$, corresponding to the gap
   which first opens as $T$ decreases towards $T_c$.

\begin{figure}[t]
\centering
\includegraphics[height=0.4\columnwidth,angle=-90]{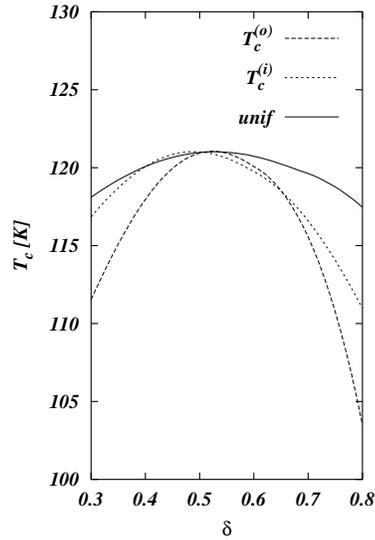}
\caption{Critical temperatures $T_c^{(i,o)}$ for inner and outer
   layers in the case of an $n=3$ layered complex, as a function of
   the overall doping $\delta$.
A nonuniform hole-content distribution among inequivalent layers has
   been assumed, according to the point-charge model.
The solid curve corresponds to the uniform distribution.
A crossover from superconductivity arising from inner or outer layers
   manifests itself as a kink in the doping dependence of
   $\displaystyle T_c = \max_\ell T_c^{(\ell)}$.
Redrawn after Reference~\protect\cite{Angilella:99b}.}
\label{fig:Tcl}
\end{figure}

The result of our numerical solutions of the equations for
   $T_c^{(\ell)}$ are shown in Figure~\ref{fig:Tcl}, where we have
   assumed a nonuniform hole-content distribution among inequivalent
   layers according to the point-charge model \cite{Haines:92}.
Such a model estimates the hole contents within inner and outer layers
   as a function of overall doping $\delta$ by minimizing the total
   carrier energy in the layered complex, expressed as a sum of band
   energy, plus electrostatic energy, where the charge distribution
   within a given layer is described as localized on the constituent
   ions of the unit cell \cite{Haines:92}.
From Figure~\ref{fig:Tcl}, it turns out that, as a function of overall
   doping $\delta$ (which can be varied by means of applied pressure),
   a crossover takes place between superconductivity arising from
   inner and outer layers.
These crossovers manifest themselves as \emph{kinks} in the doping
   dependence (pressure dependence) of $T_c$, as is experimentally
   observed in many layered cuprate superconductors
   \cite{Wijngaarden:99}.

\section{Effects of proximity to an electronic topological transition}
\label{sec:ETT}

Applied pressure can modify the properties of a bulk metal by
   changing the topology of its Fermi surface (FS), as was earlier
   recognized by I. M. Lifshitz \cite{Lifshitz:60}.
Lifshitz coined the term electronic topological transition (ETT) in
   order to describe the anomalies in several electronic properties at
   $T=0$ induced by a change of the connectivity number of the FS.
An ETT can be driven by several external agents, such as isotropic
   pressure, anisotropic deformation, and the introduction of
   isovalent impurities.
In all these cases, it is customary to employ a single critical
   parameter $z=\mu-\epsilon_c$, measuring the distance of the
   chemical potential $\mu$ from the critical value $\epsilon_c$,
   corresponding to the transition.
As $z\to0$, several normal state electronic properties, such as
   conductivity, specific heat, thermoelectric power, thermal
   expansion and sound absorption coefficients, exhibit an anomalous
   behaviour, characterized by the appearance of a step or a cusp-like
   $z$-dependence \cite{Varlamov:89,Blanter:94}.

\begin{figure}
\centering
\begin{minipage}[c]{0.6\columnwidth}
\vskip \columnwidth
\includegraphics[bb=537 77 68 555,clip,angle=-90,width=\columnwidth]{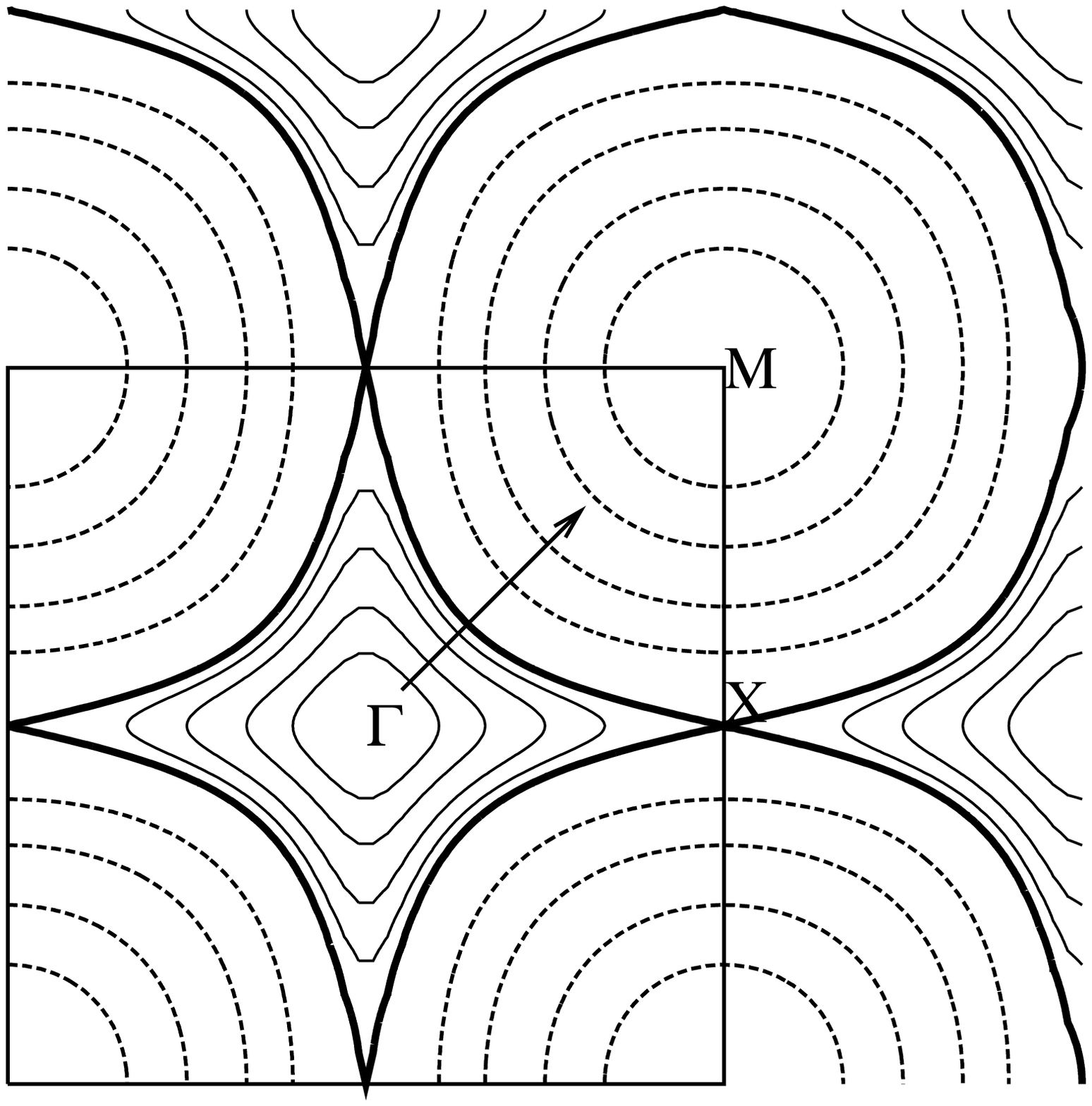}
\end{minipage}
\caption{Typical Fermi lines of a high-$T_c$ superconductor as a
   function of doping.
See text for discussion.}
\label{fig:FS}
\end{figure}

ETTs are also possible in fermion systems characterized by a low
   effective dimensionality.
This is possibly the case of the high-$T_c$ cuprates and of the
   $\kappa$-(BEDT-TTF)$_2$-X organic salts \cite{Angilella:02}.
Figure~\ref{fig:FS} displays the typical Fermi lines of a high-$T_c$ cuprate
   superconductor, such as LSCO, in the first and adjacent Brillouin
   zones ($-\pi\leq k_x , k_y \leq 2\pi$).
These correspond to $\xi_\bk = 0$ in Equation~(\ref{eq:disp}) (now
   with $t_z = 0$ and $t^\prime /t = 0.45$).
As the chemical potential $\mu$ increases from the bottom to the top
   of the band, an ETT is traversed as soon as the FS touches the
   borders of the 1BZ.
Correspondingly, the Fermi line evolves from a closed, electron-like,
   to an open, hole-like, contour, with respect to the $\Gamma$ point
   in the 1BZ.
Exactly the same kind of patterns have been recently observed
   experimentally, by means of angular-resolved photoemission
   spectroscopy (ARPES), in LSCO as a function of doping
   \cite{Ino:01}.
Close to the ETT, the band dispersion Equation~(\ref{eq:disp}) can be
   approximated by a locally hyperbolic dispersion relation,
   $\epsilon_\bk -\epsilon_c
   \approx p_1^2 / (2m_1 ) - p_2^2 /(2m_2 )$, with $p_1 = k_x$, $p_2 =
   k_y -\pi$, $\epsilon_c = -4t^\prime$, and $m_{1,2} = [2t(1\pm 2r)]^{-1}$, the fine details
   thereof being determined by the value of the next-nearest neighbours
   to nearest neighbours hopping ratio, $r=t^\prime /t$
   \cite{Alvarez:98,Angilella:01}.
In order to have a flat minimum in $\xi_\bk$ around the $\Gamma$
   point, as is observed experimentally for the majority of the
   cuprates, the condition $0<r<\frac{1}{2}$ must be fulfilled.
A universal, direct correlation between optimal doping
   $T_c^{\mathrm{max}}$ and the hopping ratio $r$ of several cuprate
   compounds has been earlier recognized by Pavarini \emph{et al.}
   \cite{Pavarini:01}, thus showing that in-plane anisotropy,
   corresponding to a relatively large value of the hopping ratio $r$,
   enhances superconductivity.

The first and foremost effect of an ETT in the spectrum of a quasi-2D
   pure electronic system is that of producing a logarithmic Van~Hove
   singularity in the density of states (DOS) each time the Fermi
   level traverses a saddle point in the dispersion relation
   \cite{Markiewicz:97,Bouvier:98,Gabovich:01}.
Due to the effect of impurities, however, such a singularity is
   smeared into a pronounced maximum with finite height
   \cite{Angilella:01}.
Moreover, two close singularities, such as those resulting from a
   pressure-induced tetragonal to orthorhombic distortion of the
   lattice in the cuprates are expected to merge into a single,
   broader maximum.

\begin{figure}
\centering
\includegraphics[height=0.7\columnwidth,angle=-90]{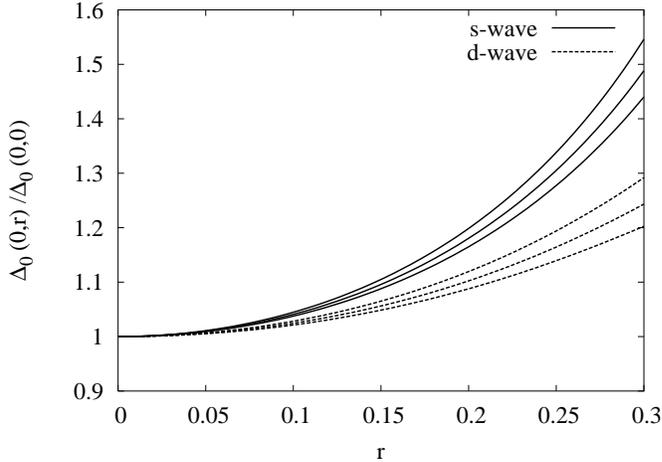}
\caption{Normalized gap amplitude $\Delta_0 (z=0,r)/\Delta_0
   (z=0,r=0)$ at $T=0$, as a function of the hopping ratio $r=t^\prime
   /t$, for different electron-electron couplings ($\lambda = 0.9 -
   1.1$, increasing from bottom to top).
Solid lines refer to the $s$-wave case, while dashed lines refer to
   the $d$-wave case.
One can recognize the direct correlation between $T_c^{\mathrm{max}}
   \propto \Delta_0 (z=0)$ and $r$, as argued by Pavarini \emph{et
   al.} \protect\cite{Pavarini:01}.
Redrawn after Reference~\protect\cite{Angilella:01}.
}
\label{fig:Tcr}
\end{figure}

As a consequence of the presence of an ETT in the electronic spectrum,
   the gap magnitude $\Delta_0$ at $T=0$ is characterized by a
   nonmonotonic dependence on the critical parameter $z$.
Within a mean-field approach to a model Hamiltonian like
   Equation~(\ref{eq:Hamiltonian}), we derived such a result, both
   analitically \cite{Angilella:01} and numerically
   \cite{Angilella:02d}, both in the $s$- and in the $d$-wave case.
In view of the fact that $T_c \propto \Delta_0$, as in any mean-field
   theory, such a finding is in agreement with the phenomenology of
   the high-$T_c$ cuprates \cite{Zhang:93,Wijngaarden:99}.
We could also estimate the maximum gap at $T=0$, which is expected to
   scale with $T_c^{\mathrm{max}}$, as the value of $\Delta_0$ close
   to the ETT ($z=0$).
Our analytical results for $\Delta_0 (z=0)$ as a function of the
   hopping ratio $r$ are shown in Figure~\ref{fig:Tcr}.
One can recognize the direct correlation between $T_c^{\mathrm{max}}
   \propto \Delta_0 (z=0)$ and $r$, as argued by Pavarini \emph{et
   al.} \protect\cite{Pavarini:01}.
Such a correlation results to be universal, in the sense that it does
   not depend on the symmetry of the order parameter and on the
   coupling strength, but rather looks like a generic consequence of
   the proximity to an ETT at $T=0$.

The ``intrinsic'' effect of applied pressure on $T_c$ has been
   probably singled out in the epitaxial strain experiments of Locquet
   \emph{et al.} on LSCO thin films \cite{Locquet:98}.
These workers investigated the effect of tensile and compressive
   epitaxial strains on the transport properties of
   La$_{2-x}$Sr$_x$CuO$_4$ (LSCO) thin films, which were epitaxially
   grown on a SrLaAlO$_4$ (SLAO) substrate, characterized by an in-plane
   lattice spacing slightly smaller than that of LSCO
   ($a_{\mathrm{LSCO}} = 3.777$~\AA, $a_{\mathrm{SLAO}} = 3.755$~\AA).
It was shown that such an in-plane compressive strain increased $T_c$ up
   to $49$~K in slightly underdoped LSCO ($x=0.11$).
This important finding is so much more interesting, since in the LSCO
   compound the hole concentration is mostly determined by the Sr
   content, together with a small oxygen non-stoichiometry, and seems
   to be little depending on pressure \cite{Murayama:91}.
In other words, the condition $dn/dP=0$ seems to apply to
   Equation~(\ref{eq:Schilling}), so that the total pressure
   derivative of $T_c$ can be attributed only to ``intrinsic''
   effects.

Within a Ginzburg-Landau phenomenological model, we have discussed the
   dependence of $T_c$ on the microscopic strains for a tetragonal
   cuprate superconductor \cite{Angilella:02d}.
For small strains, we found a monotonic increase of $T_c$ with
   increasing size of the CuO octahedron, whereas we found a
   nonmonotonic dependence of $T_c$ on either apical or in-plane
   microstrains, as well as on hydrostatic pressure, in agreement with
   the phenomenology of the high-$T_c$ cuprates \cite{Angilella:02d}.
Moreover, under epitaxial strain $\varepsilon_{\mathrm{epi}}$, we
   found a monotonically decreasing $T_c$ in the experimentally
   accessible range $-0.006 \leq \varepsilon_{\mathrm{epi}} \leq
   0.006$, with a sharp maximum just below the lower bound of the
   range \cite{Angilella:02d}, in good qualitative and quantitative agreement with the
   experimental results for epitaxially strained LSCO
   \cite{Locquet:98}.

\begin{figure}
\begin{center}
\begin{minipage}[c]{0.3\columnwidth}
\begin{center}
\includegraphics[bb=76 66 546 511,clip,width=\columnwidth]{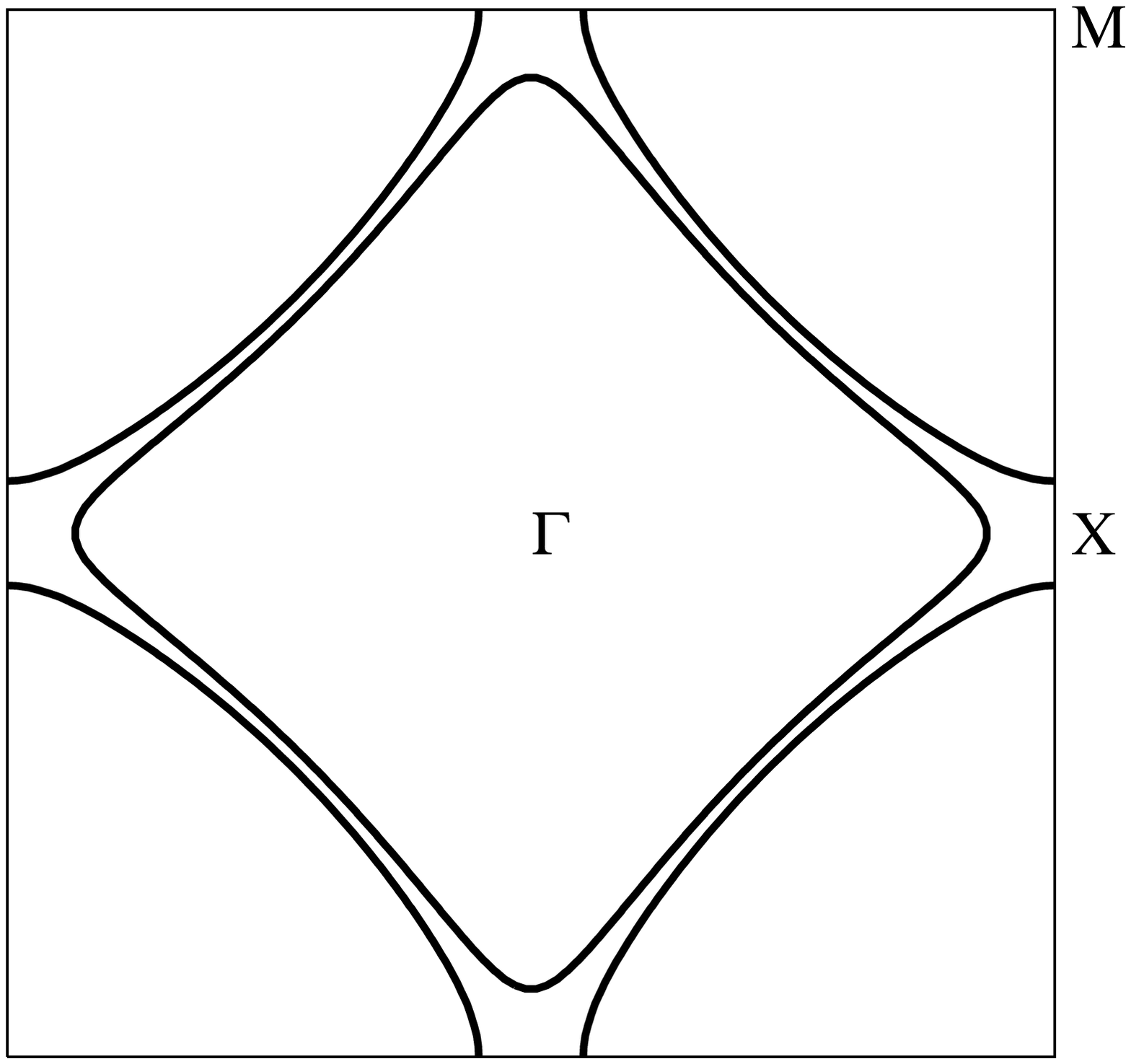}
\includegraphics[bb=76 66 546 511,clip,width=\columnwidth]{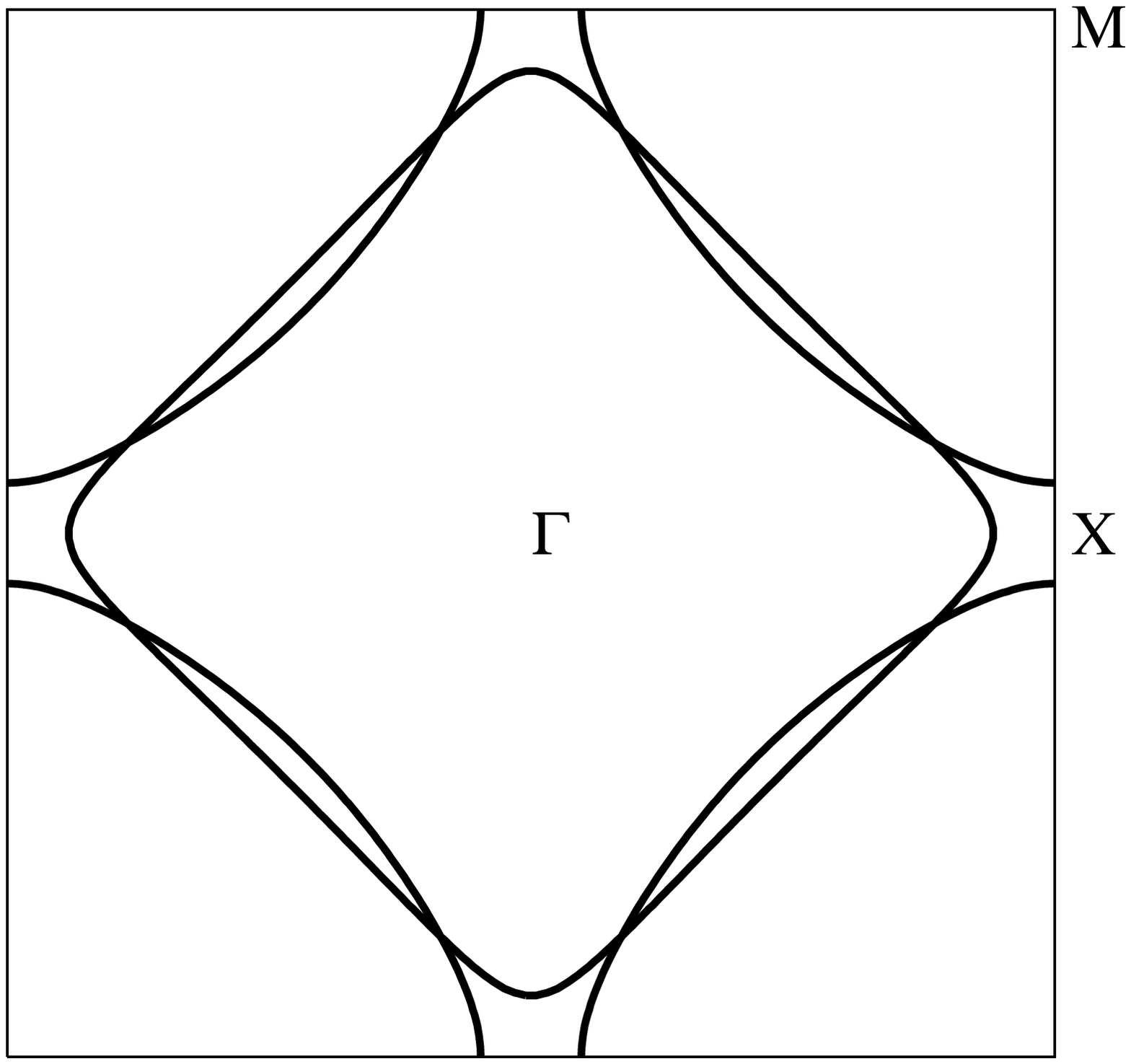}
\end{center}
\end{minipage}
\begin{minipage}[c]{0.65\columnwidth}
\includegraphics[height=\columnwidth,angle=-90]{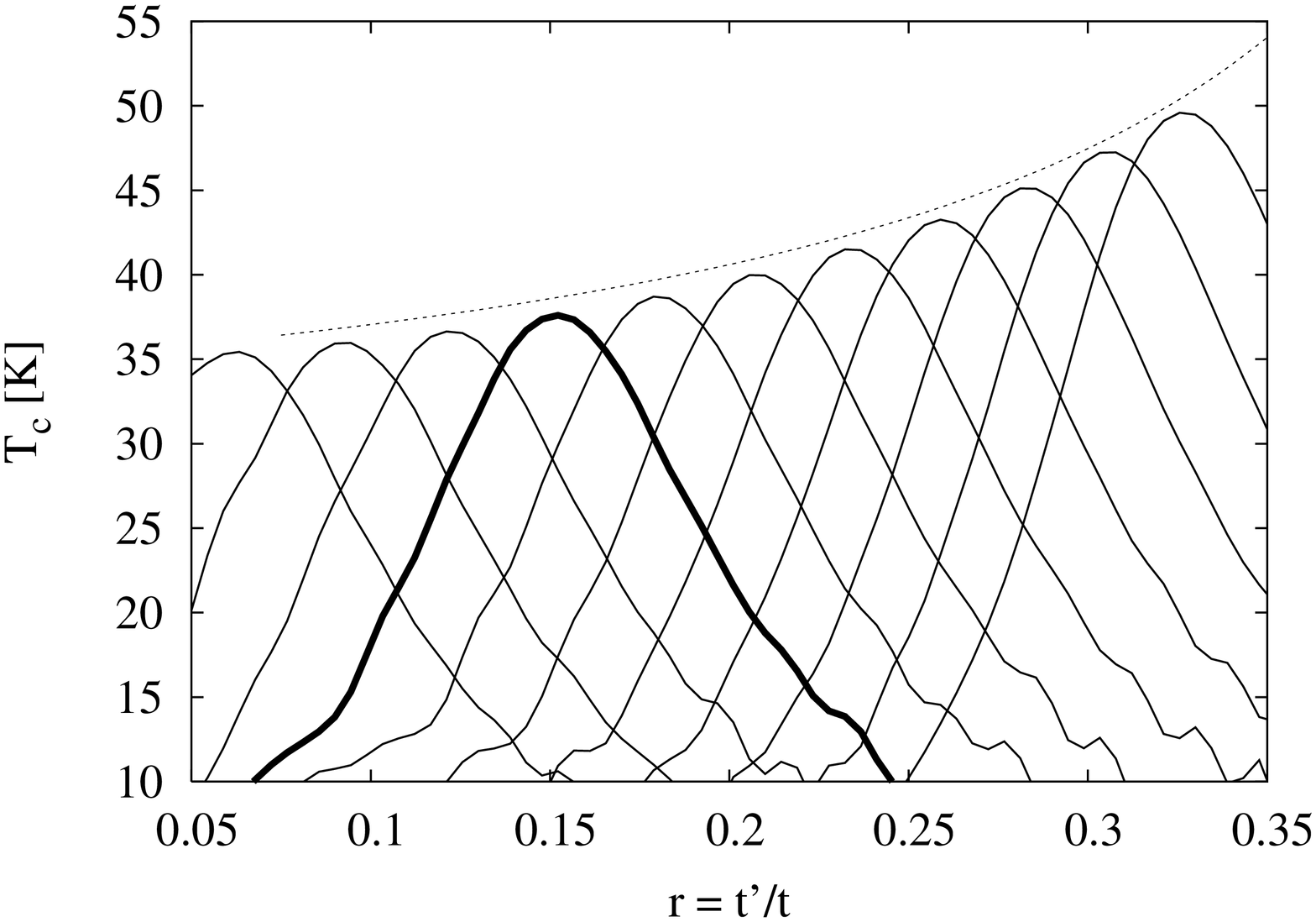}
\end{minipage}
\end{center}
\caption{{\sl Left:} Typical Fermi lines $\xi_\bk =0$,
   Equation~(\protect\ref{eq:disp}), at either side of an ETT.
{\sl Left, top panel:} Hole-content driven ETT, at constant hopping
   ratio ($r=0.182$).
{\sl Left, bottom panel:} Strain-induced ETT, at constant hole
   concentration ($\delta=0.125$, close to optimal doping).
{\sl Right:} Critical temperature $T_c$ as a function of hopping ratio
   $r=t^\prime /t$.
Along each curve, the hole concentration has been kept equal to a
   constant value ($\delta=0.05-0.3$, with $\delta=0.125$ for the
   thicker line).
The dashed line is a guide for the eye for the dependence of
   $T_c^{\mathrm{max}}$ on hopping ratio $r$.
Redrawn after Reference~\protect\cite{Angilella:02d}.
}
\label{fig:ETTr}
\end{figure}

From a microscopic point of view, a nonmonotonic strain dependence of
   the critical temperature in the high-$T_c$ cuprates may be
   interpreted as due to the proximity to an ETT.
In spite of being induced by a change in the hole concentration, as is
   usually assumed, here we considered a change of topology of the
   Fermi line as due to a strain-induced variation of the band
   parameters, such as the next-nearest to nearest neighbours hopping
   ratio $r=t^\prime /t$, at constant doping.
Figure~\ref{fig:ETTr} shows how an ETT may be driven by a change in
   $r$, at constant hole doping $\delta$.
One recognizes a nonmonotonic dependence of $T_c$ on strain, through
   the hopping ratio $r$, the maximum in $T_c$ being attained close to
   the ETT.
One also recovers the direct correlation between $T_c^{\mathrm{max}}$
   and the hopping ratio $r$
   \cite{Pavarini:01,Angilella:01,Angilella:02d}.

We have also studied the effects of the proximity to an ETT on several
   normal-state properties of the high-$T_c$ cuprates
   \cite{Angilella:01,Angilella:03g}.
In the normal state just above $T_c$, the effect of the fluctuations
   can contribute to a better understanding of the unconventional
   properties of these materials \cite{Timusk:99,Larkin:02}.
In particular, most experimental findings can be interpreted as an
   effect of precursor pairing above $T_c$ in the pseudogap region
   (see, \emph{e.g.,} \cite{Angilella:00a}, and references therein).

In the case of several transport properties in the normal state, such
   as the Ettinghausen effect, the Nernst effect, the thermopower, the
   electrical conductivity, and the Hall conductivity, the effect of
   fluctuations above $T_c$ can be embodied in the relaxation rate
   $\gamma$ of the order parameter within the time-dependent
   Ginzburg-Landau (TDGL) theory for a layered superconductor
   \cite{Larkin:02,Varlamov:99}.
In particular, a nonzero imaginary part of such a quantity can be
   related to an electron-hole asymmetry of the band structure
   \cite{Fukuyama:71,Aronov:95,Nagaoka:98}, as is the case with a
   nonzero hopping ratio $r=t^\prime /t$ in Equation~(\ref{eq:disp}).
We have evaluated $\Im\gamma$ close to an ETT, both numerically and analitically, in
   the relevant limits, as a function of the critical parameter $z$
   and temperature \cite{Angilella:03g}.
We found that $\Im\gamma$ is a sign-changing function of $z$, with two
   peaks occurring on both sides of the ETT, whose absolute value
   increases with increasing $r$.

On one hand, the anomalous behaviour of $\Im\gamma$ close to an ETT is
   a fingerprint of quantum criticality at $T=0$, and is suggestive of
   the fact that an ETT at $T=0$ may be the source of quantum critical
   fluctuations at the basis of the phase diagram of the high-$T_c$
   cuprates
   \cite{Onufrieva:96,Onufrieva:99a,Onufrieva:99b,Onufrieva:00,Bianconi:00}.
On the other hand, the direct correlation of
   $|\Im\gamma_{\mathrm{peaks}}|$ on the hopping ratio $r$ is in
   agreement with the universal behaviour of the fluctuation Hall
   effect in the cuprates \cite{Angilella:03g}, and is again an
   indication that in-plane anisotropy enhances superconductivity.

\section{Summary and concluding remarks}
\label{sec:summary}

We have reviewed our results concerning the pressure dependence of the
   critical temperature $T_c$ in the high-$T_c$ cuprate
   superconductors.
Within a mean-field approach to model Hamiltonians, we have evaluated
   $T_c$ as a function of hole content for a bilayered cuprate, and as
   a function of hole-content distribution for a multilayered cuprate,
   characterized by inequivalent layers.
In the case of a bilayered cuprate, by comparison with available
   experimental data for Bi2212 under pressure ($P=0-1.6$~GPa), we
   were able to provide a phenomenological estimate of the
   ``intrinsic'' contribution $\partial T_c /\partial P$ to the total
   pressure derivative of $T_c$, $dT_c /dP$ \cite{Schilling:01}.
In the case of a multilayered cuprate, the crossover among the
   `critical temperatures' associated to superconductivity setting in
   in inequivalent layers as a function of the overall hole content is
   in agreement with the observed kinks in $T_c$ as a function of
   pressure for these layered cuprates ($P=0-40$~GPa) \cite{Wijngaarden:99}.

We eventually proposed a microscopic interpretation of the observed
   pressure dependence of $T_c$, particularly of its ``intrinsic''
   dependence, in terms of the proximity to an electronic topological
   transition.
Instead of an ETT induced by a change in the hole content, as is
   usually assumed, we considered a strain-induced ETT, as is possibly
   the case in recent experiments in epitaxially strained LSCO
   \cite{Locquet:98}.
In this case, epitaxial strain may induce a change of the
   topology of the Fermi line by modifying the in-plane band
   parameters, at constant hole doping.
Our results for the dependence of $T_c$ and the anomalous size of some
   fluctuation transport properties above $T_c$ on the hopping ratio
   $r$ is in agreement with the universal behaviour of the high-$T_c$
   cuprates.
In particular, an increase of the in-plane anisotropy enhances
   superconductivity.

\ack

I gratefully acknowledge Professor R. Pucci for having introduced me to the
   fields of superconductivity and high pressure research, and for his
   continuous support over the years.
I also thank Professors F. Siringo, N. H. March, A. A. Varlamov,
   A. Sudb\o{} for stimulating collaborations and for sharing their invaluable
   physical insight with me.
I also acknowledge Dr. G. Piccitto for many relevant discussions and support.
Some of the results reviewed in the present Lecture have been obtained in
   collaboration with
G. Balestrino,
P. Cermelli,
F. Onufrieva,
E. Piegari,
P. Podio-Guidugli,
R. Pucci,
F. Siringo,
A. A. Varlamov.
This EHPRG Award Lecture is dedicated to my family.

\section*{References}

\bibliographystyle{mprsty}
\bibliography{a,b,c,d,e,f,g,h,i,j,k,l,m,n,o,p,q,r,s,t,u,v,w,x,y,z,zzproceedings,Angilella,notes}

\begin{thebibliography}{10}

\bibitem{Ashcroft:02}
N.~W. Ashcroft, Nature {\bf 419},  569  (2002).

\bibitem{Shimizu:01}
K. Shimizu, T. Kimura, S. Furomoto, K. Takeda, K. Kontani, Y. Onuki, and K.
  Amaya, Nature {\bf 412},  316  (2001).

\bibitem{Shimizu:02a}
K. Shimizu, H. Ishikawa, D. Takao, T. Yagi, and K. Amaya, Nature {\bf 419},
  597  (2002).

\bibitem{Christensen:01a}
N.~E. Christensen and D.~L. Novikov, Phys. Rev. Lett. {\bf 86},  1861  (2001).

\bibitem{Siringo:97}
F. Siringo, R. Pucci, and G.~G.~N. Angilella, High Press. Research {\bf 15},
  255  (1997), preprint {\tt cond-mat/9512011}.

\bibitem{Neaton:99}
J.~B. Neaton and N.~W. Ashcroft, Nature {\bf 400},  141  (1999).

\bibitem{Hanfland:99}
M. Hanfland, I. Loa, K. Syassen, U. Schwarz, and K. Takemura, Sol. State
  Commun. {\bf 112},  123  (1999).

\bibitem{Hanfland:00}
M. Hanfland, K. Syassen, N.~E. Christensen, and D.~L. Novikov, Nature {\bf
  408},  174  (2000).

\bibitem{Christensen:01}
N.~E. Christensen and D.~L. Novikov, Sol. State Commun. {\bf 119},  477
  (2001).

\bibitem{Angilella:02e}
G.~G.~N. Angilella, F. Siringo, and R. Pucci, Eur. Phys. J. B {\bf 32},  323
  (2003).

\bibitem{Anderson:97}
P.~W. Anderson, {\em The Theory of Superconductivity in the High-{$T_c$}
  Cuprates} (Princeton University Press, Princeton NJ, 1997).

\bibitem{Schilling:01}
J.~S. Schilling,  in {\em Frontiers of high pressure research {II}: Application
  of high pressure to low-dimensional novel electronic materials}, Vol.~48 of
  {\em NATO Science Series}, edited by H.~D. Hochheimer, B. Kuchta, P.~K.
  Dorhout, and J.~L. Yarger (Kluwer, Dordrecht, 2001).

\bibitem{MgB2}
A negative pressure derivative $dT_c /dP$ seems to be characteristic also of
  the non-cuprate compound MgB$_2$, which has been recently found to be
  superconducting with $T_c \sim 40$~K. [See B. Lorenz, R. L. Meng, and C. W.
  Chu, Phys. Rev. B {\bf 64}, 012507 (2001); T. Tomita, J. J. Hamlin, J. S.
  Schilling, D. G. Hinks, and J. D. Jorgensen, Phys. Rev. B {\bf 64}, 092505
  (2001)]. Such a result would be consistent with a phonon-mediated
  superconducting mechanism in this material.

\bibitem{Wijngaarden:99}
R.~J. Wijngaarden, D.~T. Jover, and R. Griessen, Physica B {\bf 265},  128
  (1999).

\bibitem{Zhang:93}
H. Zhang and H. Sato, Phys. Rev. Lett. {\bf 70},  1697  (1993).

\bibitem{Angilella:96}
G.~G.~N. Angilella, R. Pucci, and F. Siringo, Phys. Rev. B {\bf 54},  15471
  (1996).

\bibitem{Angilella:99b}
G.~G.~N. Angilella and R. Pucci, Physica B {\bf 265},  136  (1999).

\bibitem{Chakravarty:93}
S. Chakravarty, A. Sudb{\o}, P.~W. Anderson, and S. Strong, Science {\bf 261},
  337  (1993).

\bibitem{Angilella:99}
G.~G.~N. Angilella, R. Pucci, F. Siringo, and A. Sudb{\o}, Phys. Rev. B {\bf
  59},  1339  (1999).

\bibitem{Angilella:00}
G.~G.~N. Angilella, A. Sudb{\o}, and R. Pucci, Eur. Phys. J. B {\bf 15},  269
  (2000).

\bibitem{Angilella:01}
G.~G.~N. Angilella, E. Piegari, and A.~A. Varlamov, Phys. Rev. B {\bf 66},
  014501  (2002).

\bibitem{Angilella:02d}
G.~G.~N. Angilella, G. Balestrino, P. Cermelli, P. {Podio-Guidugli}, and A.~A.
  Varlamov, Eur. Phys. J. B {\bf 26},  67  (2002).

\bibitem{Angilella:03g}
G.~G.~N. Angilella, R. Pucci, A.~A. Varlamov, and F. Onufrieva, Phys. Rev. B
  {\bf 67},  134525  (2003).

\bibitem{Angilella:00b}
G.~G.~N. Angilella, N.~H. March, and R. Pucci, Phys. Rev. B {\bf 62},  13919
  (2000).

\bibitem{Angilella:01a}
G.~G.~N. Angilella, N.~H. March, and R. Pucci, Phys. Rev. B {\bf 65},  092509
  (2002).

\bibitem{Angilella:02g}
G.~G.~N. Angilella, F.~E. Leys, N.~H. March, and R. Pucci, submitted to J.
  Phys. Chem. Solids {\bf ...},  ...  (2003).

\bibitem{Huang:93}
T. Huang, M. Itoh, J. Yu, Y. Inaguma, and T. Nakamura, Phys. Rev. B {\bf 48},
  7712  (1993).

\bibitem{Leggett:98}
A.~J. Leggett, Science {\bf 279},  1157  (1998).

\bibitem{Moler:98}
K.~A. Moler, J.~R. Kirtley, D.~G. Hinks, T.~W. Li, and M. Xu, Science {\bf
  279},  1193  (1998).

\bibitem{Anderson:98}
P.~W. Anderson, Science {\bf 279},  1196  (1998).

\bibitem{Haines:92}
E.~M. Haines and J.~L. Tallon, Phys. Rev. B {\bf 45},  3172  (1992).

\bibitem{Lifshitz:60}
I.~M. Lifshitz, Sov. Phys. JETP {\bf 11},  1130  (1960), [Zh. Eksp. Teor. Fiz.
  {\bf 38}, 1569 (1960)].

\bibitem{Varlamov:89}
A.~A. Varlamov, V.~S. Egorov, and A.~V. Pantsulaya, Adv. Phys. {\bf 38},  469
  (1989).

\bibitem{Blanter:94}
{Ya. M. Blanter}, M.~I. Kaganov, A.~V. Pantsulaya, and A.~A. Varlamov, Phys.
  Rep. {\bf 245},  159  (1994).

\bibitem{Angilella:02}
G.~G.~N. Angilella, E. Piegari, R. Pucci, and A.~A. Varlamov,  in {\em
  Frontiers of high pressure research {II}: Application of high pressure to
  low-dimensional novel electronic materials}, Vol.~48 of {\em NATO Science
  Series}, edited by H.~D. Hochheimer, B. Kuchta, P.~K. Dorhout, and J.~L.
  Yarger (Kluwer, Dordrecht, 2001).

\bibitem{Ino:01}
A. Ino, C. Kim, M. Nakamura, T. Yoshida, T. Mizokawa, Z. Shen, A. Fujimori, T.
  Kakeshita, H. Eisaki, and S. Uchida, Phys. Rev. B {\bf 65},  094504  (2002).

\bibitem{Alvarez:98}
J.~V. Alvarez and J. Gonz{\'a}lez, Europhys. Lett. {\bf 44},  641  (1998).

\bibitem{Pavarini:01}
E. Pavarini, I. Dasgupta, T. {Saha-Dasgupta}, O. Jepsen, and O.~K. Andersen,
  Phys. Rev. Lett. {\bf 87},  047003  (2001).

\bibitem{Markiewicz:97}
R.~S. Markiewicz, J. Phys. Chem. Solids {\bf 58},  1179  (1997).

\bibitem{Bouvier:98}
J. Bouvier and J. Bok,  in {\em The Gap Symmetry and Fluctuations in
  High-{$T_c$} Superconductors}, edited by J. Bok (Plenum Press, New York,
  1998), p.\ 37.

\bibitem{Gabovich:01}
A.~M. Gabovich, A.~I. Voitenko, J.~F. Annett, and M. Ausloos, Supercond. Sci.
  Technol. {\bf 14},  R1  (2001).

\bibitem{Locquet:98}
J. Locquet, J. Perret, J. Fompeyrine, E. Machler, J.~W. Seo, and G. {Van
  Tendeloo}, Nature {\bf 394},  453  (1998).

\bibitem{Murayama:91}
C. Murayama, Y. Iye, T. Enomoto, N. Mori, Y. Yamada, T. Matsumoto, Y. Kubo, Y.
  Shimakawa, and T. Manako, Physica C {\bf 183},  277  (1991).

\bibitem{Timusk:99}
T. Timusk and B. Statt, Rep. Prog. Phys. {\bf 62},  61  (1999).

\bibitem{Larkin:02}
A. Larkin and A.~A. Varlamov,  in {\em Handbook on Superconductivity:
  Conventional and Unconventional Superconductors}, edited by K. Bennemann and
  J.~B. Ketterson (Springer Verlag, Berlin, 2002).

\bibitem{Angilella:00a}
G.~G.~N. Angilella, N.~H. March, and R. Pucci, Phys. Chem. Liquids {\bf 38},
  615  (2000).

\bibitem{Varlamov:99}
A.~A. Varlamov, G. Balestrino, E. Milani, and D.~V. Livanov, Adv. Phys. {\bf
  48},  655  (1999).

\bibitem{Fukuyama:71}
H. Fukuyama, H. Ebisawa, and T. Tsuzuki, Prog. Theor. Phys. {\bf 46},  1028
  (1971).

\bibitem{Aronov:95}
A.~G. Aronov, S. Hikami, and A.~I. Larkin, Phys. Rev. B {\bf 51},  3880
  (1995).

\bibitem{Nagaoka:98}
T. Nagaoka, Y. Matsuda, H. Obara, A. Sawa, T. Terashima, I. Chong, M. Takano,
  and M. Suzuki, Phys. Rev. Lett. {\bf 80},  3594  (1998).

\bibitem{Onufrieva:96}
F. Onufrieva, S. Petit, and Y. Sidis, Phys. Rev. B {\bf 54},  12464  (1996).

\bibitem{Onufrieva:99a}
F. Onufrieva, P. Pfeuty, and M. Kiselev, Phys. Rev. Lett. {\bf 82},  2370
  (1999).

\bibitem{Onufrieva:99b}
F. Onufrieva and P. Pfeuty, Phys. Rev. Lett. {\bf 82},  3136  (1999), [Phys.
  Rev. Lett. {\bf 83}, 1271 (1999)].

\bibitem{Onufrieva:00}
F. Onufrieva and P. Pfeuty, Phys. Rev. B {\bf 61},  799  (2000).

\bibitem{Bianconi:00}
A. Bianconi, G. Bianconi, S. Caprara, D. {Di Castro}, H. Oyanagi, and N.~L.
  Saini, J. Phys.: Condens. Matter {\bf 12},  10655  (2000).

\end{thebibliography}

\end{document}